\documentclass[twocolumn,showpacs,amsmath,amssymb]{revtex4}

\usepackage{graphicx}% Include figure files
\usepackage{dcolumn}% Align table columns on decimal point
\usepackage{bm}% bold math
\usepackage{hyperref}

\begin{document}

\title{First-principles thermodynamics of transition metals: 
W, NiAl, PdTi} 
\author{ Graeme J. Ackland, Xiangyang Huang, and Karin M. Rabe}
\affiliation{
Department of Physics and Astronomy, Rutgers University, 
Piscataway, New Jersey 08854-8019}

\date{\today}
\begin{abstract}

We apply the pseudopotential density functional perturbation 
theory approach along with the quasiharmonic approximation to calculate 
the thermal expansion of tungsten and 
two important metallic alloys, NiAl and PdTi.  
We derive the theory for anisotropic crystal structures and test the approximation that 
the anisotropic effects of thermal expansion are equivalent to 
negative pressure - this simplifies the calculation enormously for 
complex structures.  Throughout, we
find excellent agreement with experimental results.

\end{abstract}


\maketitle

\section{Introduction}

First-principles calculations have established an excellent record for
describing the ground state properties of materials at zero
temperature over almost twenty years.  More recently, the full
description of the electronic structure allows for accurate
calculation of phonon frequencies\cite{baroni} and the thermodynamic
properties derivable from them\cite{freeE}.  These methods have been
applied to insulating systems and elemental metals.  In this paper, we
extend the methodology to investigate transition metals and 
compounds that have a low symmetry or phonon anomalies.
Previous authors have looked at elemental metals Al, Li,
Na\cite{quong}, Ag\cite{xie}, Cu\cite{narasimhan}, and W\cite{debernardi} 
and found that the quasiharmonic
approximation holds most of the way to melting.  We extend this work
to alloys, which introduces some new features.  As our prototype
materials, we study W, NiAl and PdTi, which have many similarities but
each of which has aspects of peculiar interest.

NiAl is a high  
melting point alloy which finds applications in aerospace and 
turbine blade design.  Its electronic structure is characterized 
by Kohn anomalies in the phonon spectrum\cite{huang:nial}, arising from
a Fermi surface nesting effect.  A huge amount of work has been done on 
the system using empirical potentials\cite{ackland:mrs,gao,mishin:2002,mishin:1997,yan}, 
which cannot describe the Kohn 
anomaly, and it is interesting to examine whether this neglect has 
serious structural consequences.   Crystallographically, NiAl has the cubic 
$B2$ (CsCl) crystal structure which is stable at all temperatures.  This 
high symmetry simplifies the electronic structure calculation enormously.

PdTi provides a sharp contrast to NiAl.  It is a high temperature
shape memory alloy, isostructural with NiAl at high temperatures in
the $B2$ phase, but undergoing a martensitic phase transition to the
tetragonal $B19$ phase at 810 K and predicted to undergo a further
transition to the monoclinic $B19'$ at very low 
temperatures\cite{huang:2002b}.  
The dynamically stabilized high temperature phase and low symmetry 
low temperature phase each provide a challenging test for theoretical 
prediction.

Tungsten is another high melting point transition metal.  In addition
to studying its own thermal expansion, we use this to investigate the
effects of anisotropic strain on the phonon contribution to the free
energy.  This enables us to make a simple test of some approximations
required in the calculation of PdTi.

The difference between the materials being studied 
is highlighted by the phonon spectra for the
$B2$ phase: NiAl and PdTi both adopt the $B2$ structure at high
temperature while tungsten is a refractory metal and has
a bcc structure at all temperature. The bcc structure can be regarded as
a special case of $B2$ where both atoms are the same. 
NiAl has phonon anomalies 
due to Fermi surface nesting, while $B2$ structure PdTi is dynamically 
stabilized: it has negative frequency phonons (i.e. mechanically unstable) 
at low temperatures.  Such a situation leads to temperature
dependent modes, which cannot be treated within the quasiharmonic
approximation and would need to be dealt with separately either in
real\cite{zhong} or reciprocal\cite{drummond} space.

The quasiharmonic approximation\cite{quasi} to the free energy assumes all phonons
can be treated as simple harmonic oscillators with a frequency dependent 
on the volume of the material.  The static lattice energy ($T=0$)
is evaluated at a range of volumes, and for each of these the phonon
dispersion relation is calculated.  The quasiharmonic contribution to 
the free energy at finite temperature is then evaluated from 
Bose-Einstein statistics of the phonons at a given volume.  From 
this the pressure can be evaluated for all conditions of $T$ and $V$, 
giving an equation of state. Alternately in the zero pressure case, 
the relationship between $T$ and $V$ (i.e. the thermal expansion) can 
be determined.  Since larger volumes typically lead to lower phonon
frequencies, the entropy is typically lowered by expansion.  Thus larger 
volume (entropy) is favoured at higher temperatures, and thermal expansion 
occurs.  

There are, of course, other contributions to thermal expansion in 
ordered alloys (antisite defect creation, vacancy creation, 
coupling of phonons to lattice parameters, 
anharmonic effects due to finite atomic displacements) and it is 
interesting to test how much of the 
observed expansion can be attributed to quasiharmonic effects.

\section{Theory}

Phonon dispersions can be evaluated using density-functional perturbation theory
(DFPT) \cite{baroni:1987} and finite displacement methods\cite{prac}. Here 
we use the former.
Previous work on isotropic equations of state for cubic materials 
covers much of the theory, here we investigate the practical 
requirements for lower symmetry materials, in particular 
what approximations can be made to made calculations tractable.

Virtually all implementations of density-functional theory (DFT) based
{\it ab initio} codes take as input the 
unit cell vectors {\bf a, b} and {\bf c} and atomic positions.  We define a 
matrix {\it B} comprised of the three unit cell vectors.

\[ B =(\textbf{a, b, c}) \]
 
The volume of the unit cell is  
\[ V=|B|=(\textbf{a}\times\textbf{b})\centerdot\textbf{c}.\]

For a cubic material, the quasiharmonic approximation uses the phonon
frequencies calculated at different volume to obtain the free energy 
as a function of two variables,

\begin{equation}
F(V,T) = E(V) - k_BT \ln Z(V)
\end{equation}

where $E(V)$ is the cold curve energy at the volume $V$ 
(i.e. the energy of a system with the atoms placed on their lattice sites), 
$k_B$ is the Boltzmann's constant and $Z$ is the vibrational
partition function.  We assume that for all temperatures of interest
the contribution of electronic excitations is negligible - generally
true for equation of states\cite{dswift}, but not for
transport properties such as heat conduction. 

From $F(V,T)$ one can obtain the equation of state of the 
single phase of the system: 

\begin{equation}
P = -\left(\frac{\delta F}{\delta V}\right )_T
\end{equation} 

and the volume thermal expansion:

\begin{equation}
\alpha_V = \frac{1}{V}\left(\frac{\delta V}{\delta T}\right)_P
\end{equation}

These can be evaluated for a given structure regardless of whether 
an alternate phase of lower Gibbs free energy exists.

For non-cubic materials, the free energy is a function of all
independent cell parameters, and the linear thermal expansion 
is a tensor quantity.  Calculating the free energy across such
a multidimensional space is impractical, so we examine 
a further approximation: that the effect of increasing temperature is
equivalent to a negative pressure\cite{foot}
At each pressure, the internal coordinates must be
re-relaxed to their equilibrium position within the unit cell, since
the phonon spectrum is only well defined by an expansion about such an
equilibrium.

Thus, we proceed as follows:

1/  Evaluate the equilibrium structure at $P=0$, by minimizing 
$U({\bf a, b, c}, \{  u_i \}) \equiv U_0 $ 
where $ u_i$ are the internal atomic coordinates.

2/  Evaluate the force constants and hence
the phonon dispersion relation for the relaxed structure, and the 
partition function $Z(V)$ for these oscillators.
$V=|B|=(\textbf{a}\times\textbf{b})\centerdot\textbf{c}$.

3/  Evaluate Gibbs free energy at a range of temperatures 
$F(V,T) = U_0 - kT \ln Z + PV$.
 
4/  Evaluate the equilibrium structure at P, by minimizing enthalpy
$U({\bf a, b, c}, \{u_i\}) + PV \equiv H_0(V) $ where $u_i$ are the internal 
atomic coordinates.

5/ Evaluate the phonon dispersion relation for this relaxed structure,
and the partition function $Z$ for these oscillators.

6/  Evaluate Gibbs free energy at fixed V for a range of temperatures 
$F(V,T) = U_0 - kT \ln Z + PV$.

7/ Repeat 4, 5, 6 for a range of positive and negative pressures.

Once a coarse grid of 
$V$-points has been generated (for thermal expansion, as few as 3 different 
pressures suffices) a finer grid can be generated without further expensive 
{\it ab initio} calculation by interpolating force constants\cite{dswift}.
A dense grid of $T$-points can easily be generated at a given $V$.  

In terms of speed and accuracy there is little to choose between  
DFPT or finite displacements: the former involves calculation 
of second derivatives and the later requires large supercells.  
We note that for finite displacements, once the eigenvectors 
are calculated at $P=0$, a single calculation of restoring forces
with all atoms displaced suffices at other volumes\cite{freeE}.

The allowed occupation states of each phonon
mode of frequency $\omega$ are $\epsilon_n(\omega)=(n+1/2) \hbar
\omega$ where $n$ is the 
the number of {\it phonons\/} populating that mode.
The canonical partition function is therefore 

\begin{equation} 
\xi
(\omega) \equiv \sum_{n=0}^\infty e^{-\beta \epsilon_n (\omega)}
=1/(2\sinh(\hbar \omega /2k_B T)).
\end{equation}
%%$k_B$ being Boltzmann's constant and $T$ the temperature.  

The total partition function is then:

\begin{equation} Z = \Pi_\omega \xi(\omega)
\end{equation}

For calculating thermal expansion under constant (zero) pressure boundary
condition, the essential quantity is the  free energy.  The mean
vibrational free energy:

\begin{equation} 
-k_BT\ln Z(V) =-k_BT\int_0^\infty g(\omega,V) \ln(\xi(\omega,T)) \, d\omega, \label{equation_Z_total} \end{equation}

introducing the phonon density of states  $g(\omega,V)$ 
to make the dependence of density of states on volume explicit.

%%Each wavevector has $3N$ corresponding frequencies. The
%%frequency density-of-states function $G(\omega)$ is defined such that
%%$G(\omega) d \omega$ is the number of frequencies between $\omega$ and
%%$\omega + d \omega$. Hence if L wavevectors are sampled, 
%%$G(\omega) d \omega/3NL$ can be thought of
%%as the probability that a randomly chosen allowed frequency will lie
%%between $\omega$ and $\omega+d \omega$. 

%%We calculate the specific frequency density-of-states function
%%$g(\omega)=G(\omega)/\left( L \sum_{n=1}^N m_n \right)$.
%%by dividing the frequency 
%%domain into bins of width $\delta
%%\omega$, where $\delta \omega$ is small. 
Phonon frequencies are evaluated at a  dense set of $q$-points 
in the first Brillouin zone using the force constants matrix obtained by fourier transform of the ``exact'' DFPT frequencies. 
The density of phonon states is calculated by integration using the 
tetrahedron method\cite{lehmann}.  
This provides the $Z(V)$ required in step 2/ above.

The structure obtained at a particular temperature can be found by 
minimizing the free energy with respect to the lattice vectors and 
internal coordinates of the atoms 

\begin{equation} \frac{d F}{d B_{ij} } = 0 ; \hspace{1cm} 
\frac{d F}{d {\bf u_k} } = 0 \end{equation}

These derivatives can be split into  cold curve ($E(B,{\bf u_k})$) 
and phonon ($F_{phon}$) contributions.

%\[ \frac{\partial E}{\partial  B_{ij} } |_{\bf u_k} + 
% \frac{\partial F_{phon}}{\partial  B_{ij} }|_{\bf u_k}  = 0 \]

%\[ \frac{\partial E}{\partial {\bf u_k} }|_{B_{ij}} 
% \frac{\partial F_{phon}}{\partial {\bf u_k} }|_{B_{ij}} = 0 \]

To make calculations for complex structures tractable, we
make the assumption that the phonon part of the 
free energy is a function of volume {\it only}, and is independent 
of  ${\bf u_k}$.  We examine this approximation in the case of 
anisotropic deformation of tungsten and find it to be good.  
Intuitively, this can be understood if the force constants vary 
linearly with separation: we might expect that volume conserving 
shears and changes in  ${\bf u_k}$ will stiffen some force constants 
and weaken others, raising some frequencies and lowering others to 
give an overall cancellation in the small strain limit.  By contrast, volume 
increases weaken all force constants and give a systematic decrease
in mean frequency.  Hence the conditions above become:

\begin{eqnarray} \frac{\partial E}{\partial  B_{ij} }|_{\bf u_k} = 0 \\ 
 \frac{\partial F_{phon}}{\partial {V} }|_{\bf u_k} = 0 \\
 \frac{\partial E}{\partial {\bf u_k} }|_V =0 
\end{eqnarray}

\section{Results}

\subsubsection{Anisotropic strain - Tungsten}

We use tungsten as a benchmark to investigate the effects of
anisotropic strain on the phonon contribution to the free energy.
Tungsten has a bcc crystal structure. The calculations were done
with the generalized gradient approximation (GGA) and the equilibrium
lattice parameter was computed to be 2.905 \AA, comparable to
the experimental result of 2.887 \AA. Phonon calculations were done
at three different volumes: $V_0$, 1.01$^3V_0$ and  1.02$^3V_0$.
At each volume, phonon dispersion was done with the bcc structure and tetragonal strained structures
with four different $c/a$ ratios, i.e.  0.96, 0.98, 1.02 and 1.04.

\begin{figure}
\includegraphics[scale=0.5]{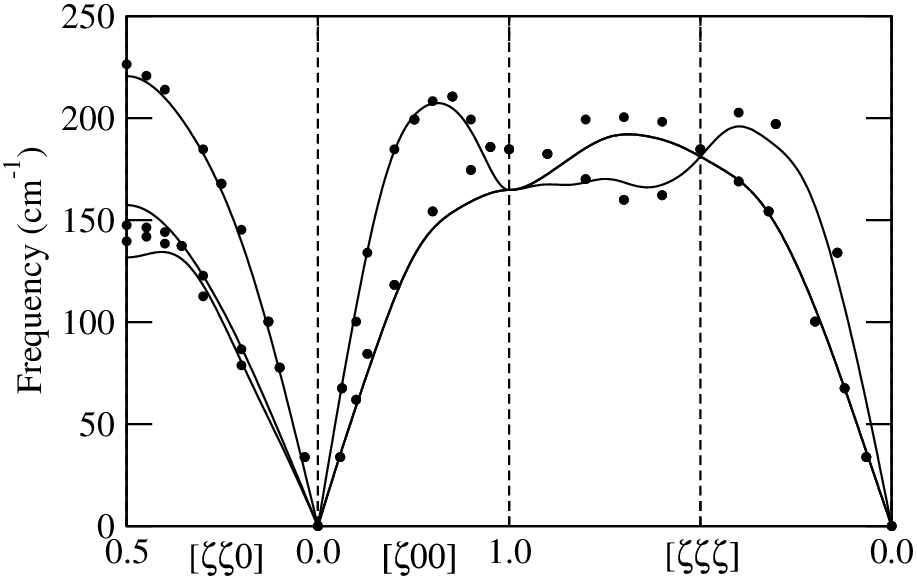}
  \caption{ \label{fig:Wphon} 
  Phonon spectra calculated for bcc W at $a_0$=2.905\AA, 
  using the {\sc PWSCF} and {\sc PHONON} code \protect\cite{SISSAcode}
  with DFPT\cite{baroni1} and
  a norm conserving pseudopotential.  The
  electronic wave functions were represented in a plane-wave basis set
  with a kinetic energy cutoff of 32 Ry.  
  The Brillouin zone (BZ) integrations were carried
  out by the Hermite-Gaussian smearing technique \cite{methfessel}
  using a $16\times 16\times 16$ {\it k}-point mesh.
  Phonon calculations are exact on meshes which are commensurate 
  with the {\it q}-point mesh: a $8\times 8\times 8$ mesh
  for tungsten. Other phonons come from dynamical matrices calculated 
  using force constants determined by Fourier transform of original 
  data\cite{baroni1}.
}
\end{figure}

\begin{figure}
\includegraphics[scale=0.5]{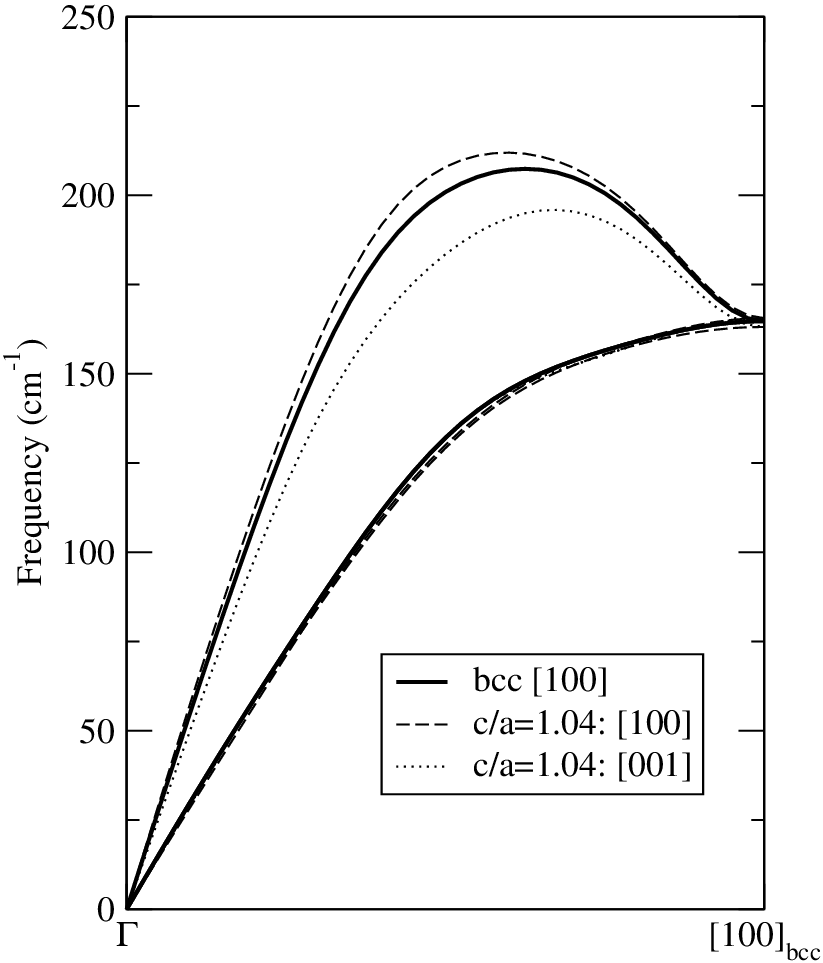}
\includegraphics[scale=0.5]{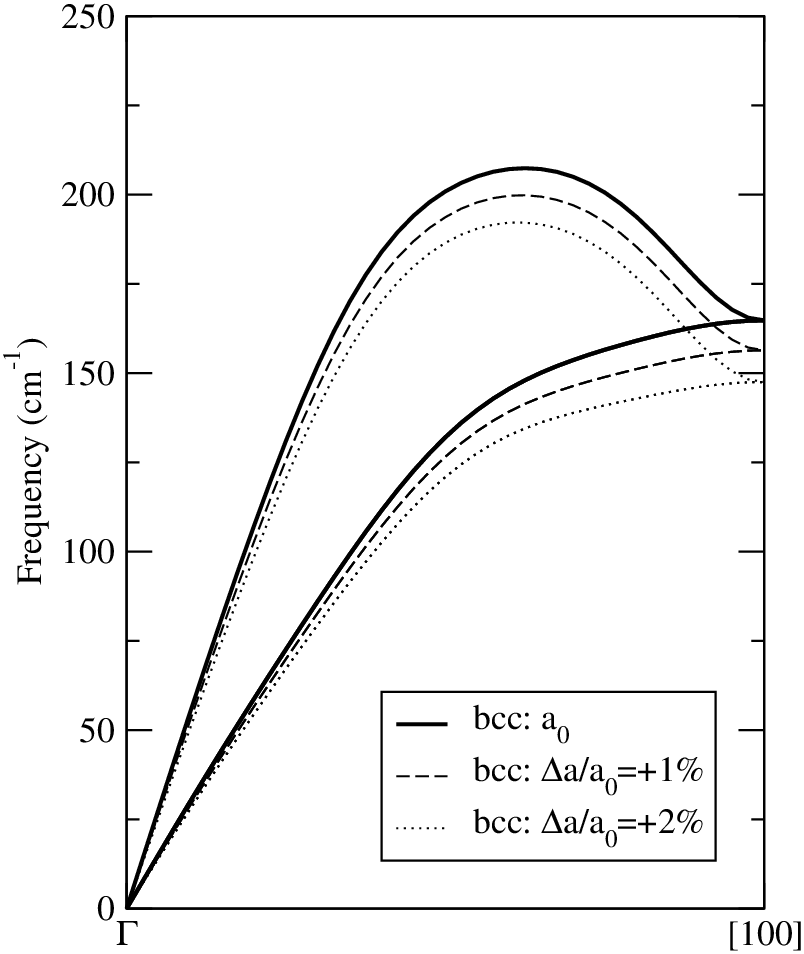}
\caption{ \label{fig:Wstrain}
  [001] and [100] branches of the tungsten dispersion relation 
  for volume-conserving strains of (a)  4\% on (001) 
  unstrained and (b) bulk strains of 1\% and 2\%.
  The key observation is that for volume conserving shear
  some frequencies increase while others decrease,  
  meanwhile for volume expansion all frequencies decrease.  
  Thus in the former case some compensation occurs
  to keep the phonon free energy constant, while in the latter case all 
  modes contribute to increase free energy.  
  Calculation details are as for Fig.~\ref{fig:Wphon}.
}
\end{figure}

We calculated phonon density of states (DOS) at 15 different structures corresponding 
to tetragonal strains ($e_{xx}$; 
$e_{yy} = e_{zz}$; $e_{xy} = e_{yz} = e_{zx} =0$).   
The phonon dispersion of bcc tungsten at equilibrium lattice parameter
along with experimental results
is shown in Fig.~\ref{fig:Wphon},
while the phonon dispersions under tetragonal
strains and a larger volume are shown in Fig.~\ref{fig:Wstrain}(a) and
Fig.~\ref{fig:Wstrain}(b), respectively.
The calculated phonon frequencies 
are in good agreement with experiment (Fig.~\ref{fig:Wphon}) and with a 
recent calculation\cite{debernardi}.
Fig.~\ref{fig:Wstrain}(a) and (b) show that compared to volumetric strain,
volume-conserving shear
strains have negligible contribution to phonon free energy.

\begin{figure}
\includegraphics[scale=0.6]{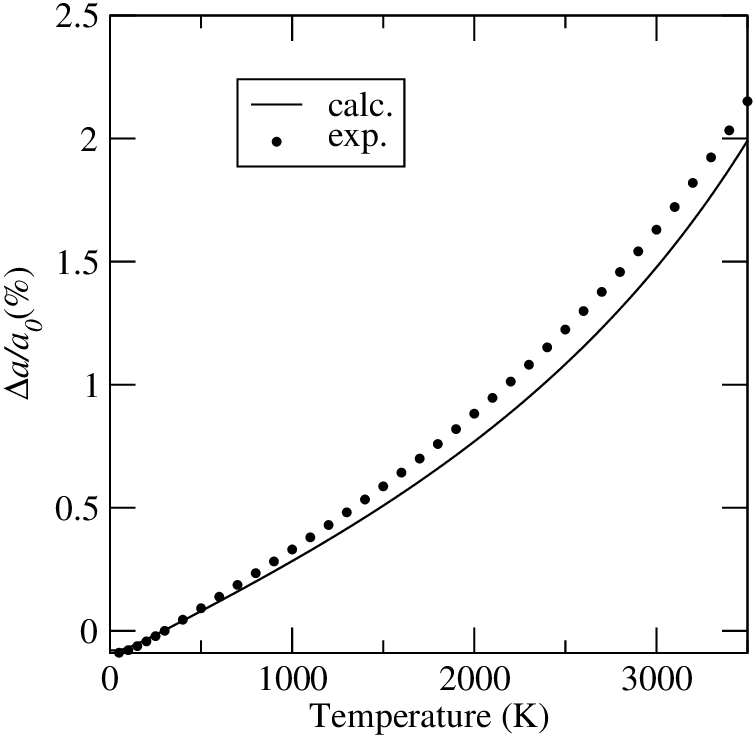}
\caption{\label{fig:Wexp}
  Calculated linear thermal expansion for tungsten, compared 
  with experimental data from Ref.\onlinecite{nash}.  $a_0$
represents the room temperature volume.
}
\end{figure}

The free energy was evaluated at each strain state.
In Fig.~\ref{fig:Wexp} we show the thermal expansion calculated using 
the three isotropic strains. The linear thermal expansion adopts the
form from Ref.~\onlinecite{amer}.  
The agreement between theoretical and experimental results is very good.

\begin{figure}
\includegraphics[scale=0.5]{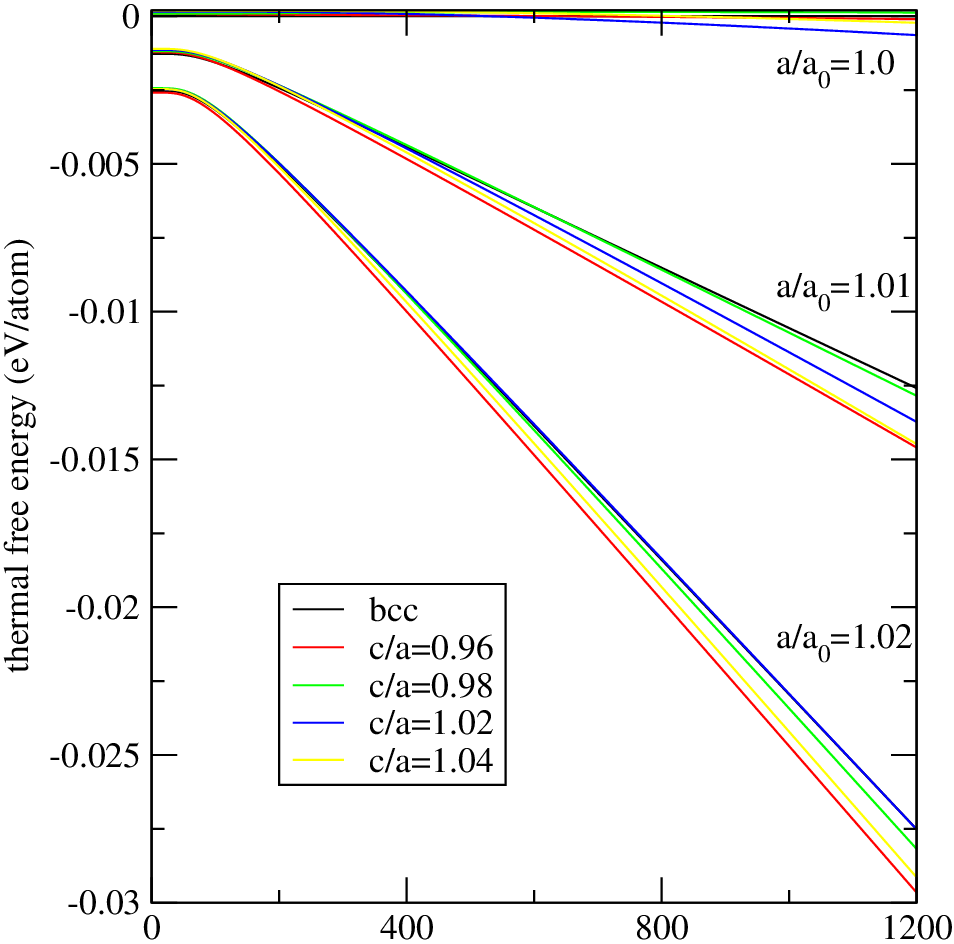}
\caption{\label{fig:Wtet}
  Graphs of tungsten phonon free 
  energy against temperature for 15 different
  strain conditions, plotted relative to $G(V)$ for $V$ at 
  the cold curve minimum.  The effect of volume strain can be seen to 
  be about ten times greater than that of shear strain.
}
\end{figure}

Fig.~\ref{fig:Wtet} shows the phonon free energy with different strain.
As expected, the phonon contribution to
the free energy is only weakly affected by
volume-conserving anisotropic strain. 
Thus it is a good approximation to neglect 
this term: a given volume-conserving strain contributes about 
5\% as much as equivalent volume strain.

\subsubsection{Anomalous phonons - NiAl}

\begin{figure}
\includegraphics[scale=0.7]{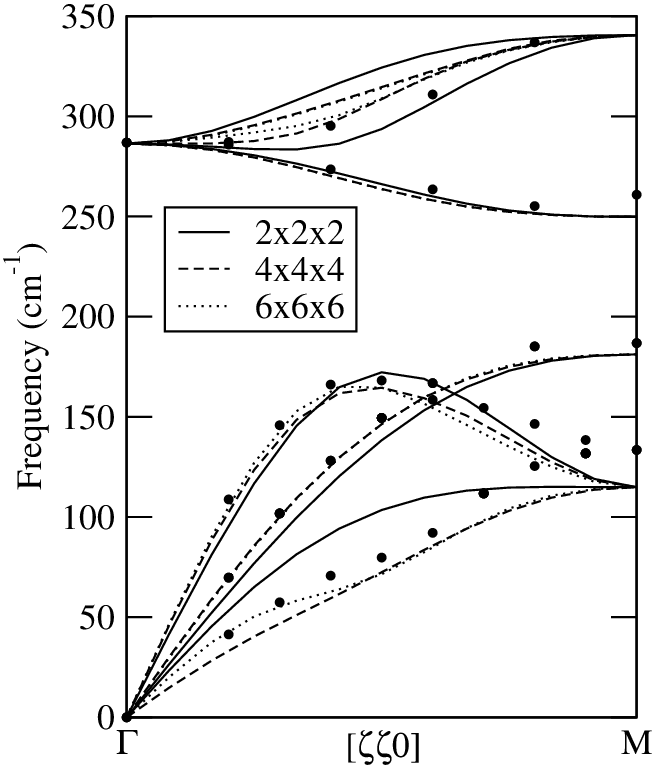}
\caption{ \label{fig-4}
  Phonon dispersion relations along the acoustic (110) branches 
  in NiAl, calculated with increasing $q$-point density.  Lines are calculated
  data, taken from force constants deduced from Fourier transforming the
  $\omega(q)$ data. Labels indicate the number of $q$-points used in the
  DFPT calculation.  Symbols indicate experimental data taken from 
  Ref.~\onlinecite{mostoller}
  for a composition of Ni$_{50}$Al$_{50}$ \label{fig:anom}
  Calculations use the {\sc PWSCF} and {\sc PHONON}\protect\cite{SISSAcode}
  code and DFPT\cite{baroni1} using an ultrasoft pseudopotentials for Ni,
  \cite{rappe}, and a
  norm conserving pseudopotential for Al.  The
  electronic wave functions were represented in a plane-wave basis set
  with a kinetic energy cutoff of 30 Ry.  
  The BZ integrations\cite{methfessel} used $12\times 12\times 12$ mesh for NiAl, 
  giving exact phonon on a $6 \times 6 \times 6$ mesh.
}
\end{figure}

The phonon spectrum of $B2$ NiAl is characterized by Kohn anomalies.  
As a consequence, to describe the phonon dispersion 
properly requires a thorough DFPT $q$-point sampling  
or (equivalently) large supercells in direct space methods.  
In Fig.~\ref{fig:anom}, we show the convergence of the result 
to experimental observation with $q$-point sampling density.

Despite the anomalies, we find that the thermodynamic properties 
are almost independent of $q$-point sampling aside from a rigid 
shift of the free energy. This does not affect the thermal expansion, 
but stabilizes the structure slightly against alternate structures
because of the increased entropy of the low frequency phonons.

\begin{figure}
\includegraphics[scale=0.45]{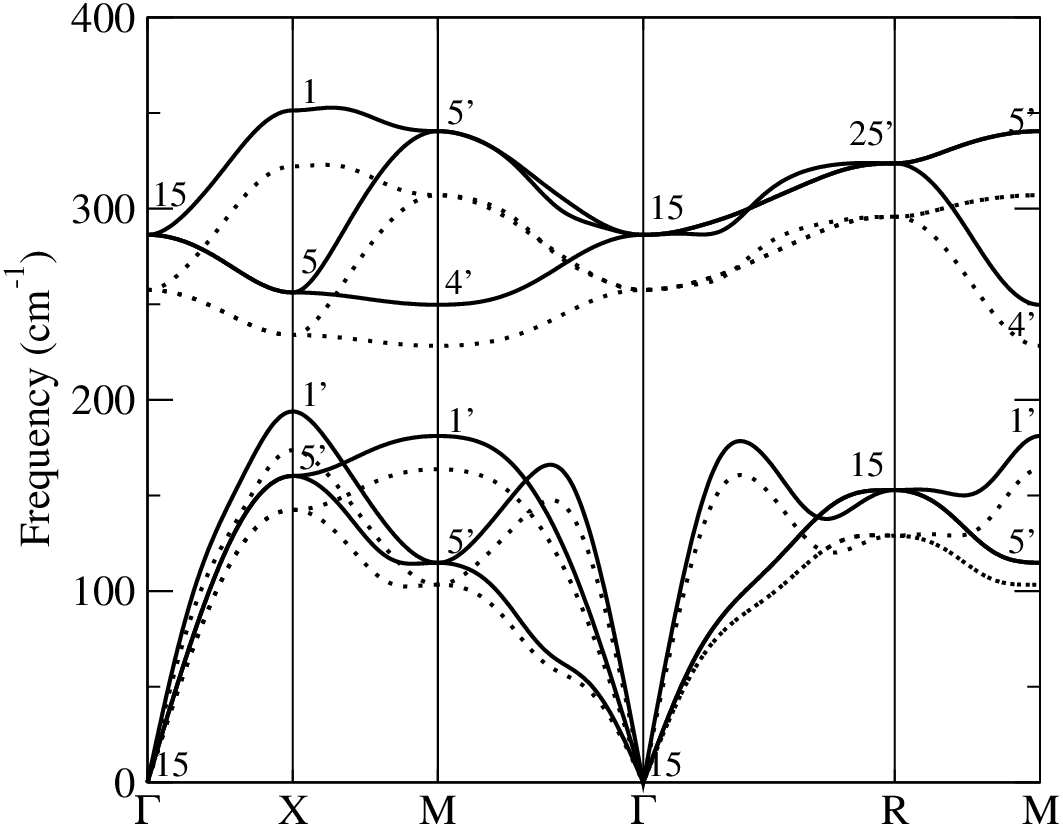}
\includegraphics[scale=0.45]{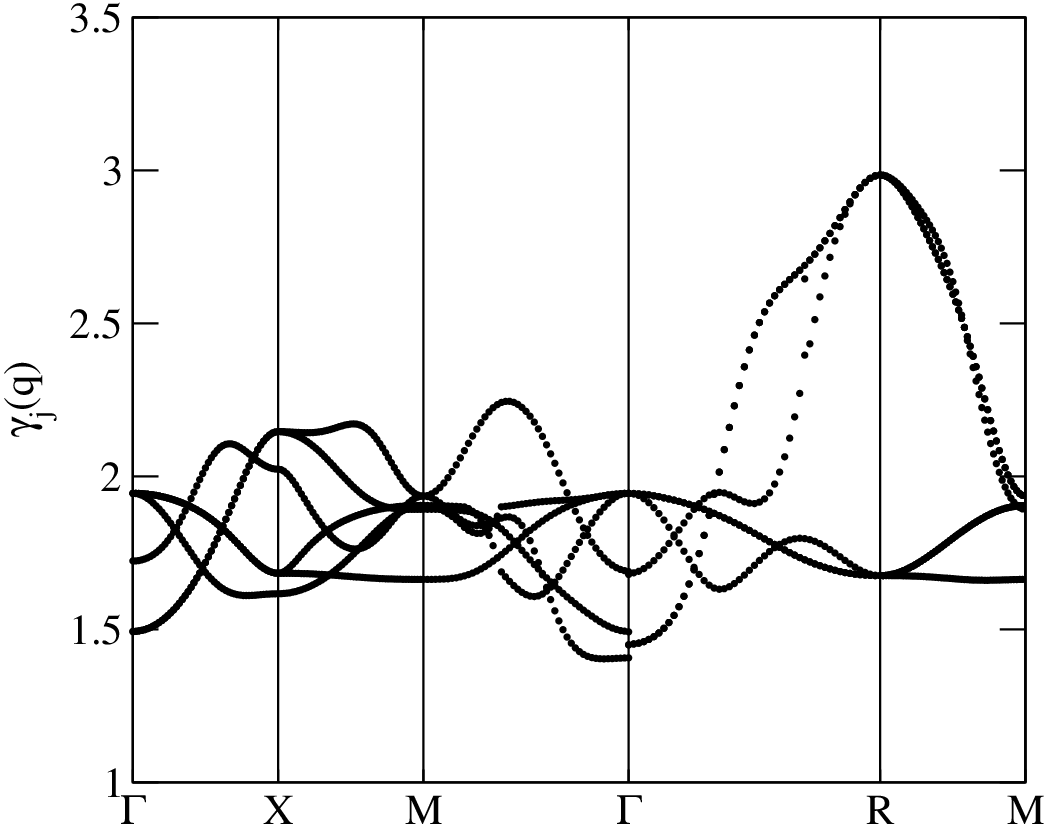}
\caption{ \label{fig:nial-phon} 
  (a) Phonon spectra calculated for $B2$ NiAl, 
  solid lines are at a lattice parameter of $a_0$=2.906\AA
  (GGA theoretical lattice constant)
  the dotted lines at a lattice of $1.02a_0$.  All phonons have higher 
  frequencies under compression.
  %, including those associated with the phonon anomaly around
  %($\frac{1}{3},\frac{1}{3},\frac{1}{3}$). 
  %This is further evidence that the anomaly is electronic in
  %origin. 
  (b) Calculated mode Gr\"uneisen parameters of NiAl along symmetry lines
  of simple cubic BZ at equilibrium volume.
}
\end{figure}

\begin{figure}
\includegraphics[scale=0.6]{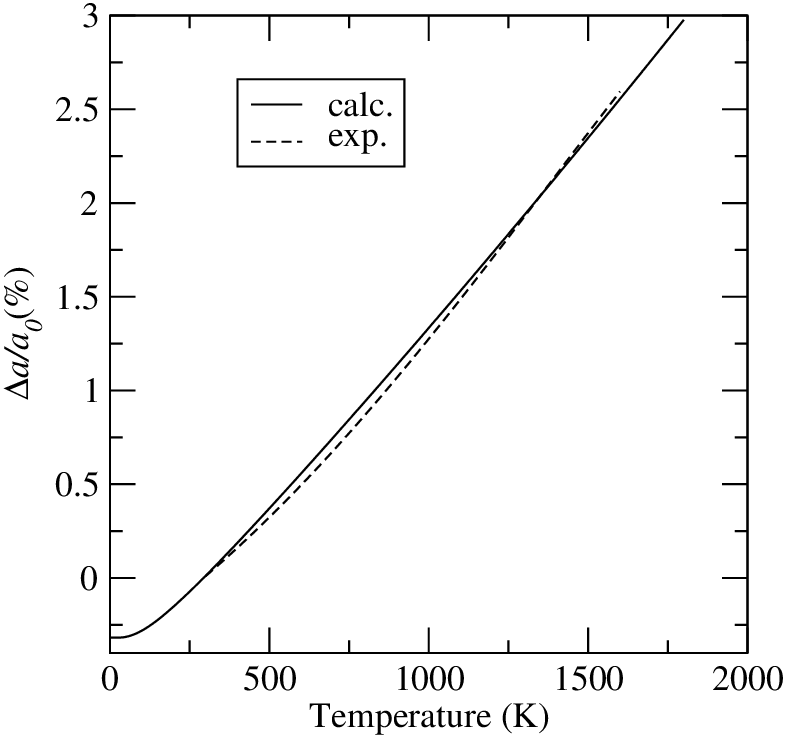}
\caption{ \label{fig:nial-exp}
  Calculated thermal expansion for NiAl, 
  compared with experimental data from Ref.~\onlinecite{nepijko}. The 
  zero of lattice parameter $a_0$ is chosen to be 
  2.905\AA in the theory and 2.886\AA in the experimental case, which correspond to 273K}
\end{figure}

The entire calculated phonon dispersion relations at two 
different volumes are shown in Fig.~\ref{fig:nial-phon}(a).
All phonon frequencies increase under compression.
Hence the mode Gr\"uneisen parameters, $\gamma_{j}(\textbf{q})$, 
shown in Fig.~\ref{fig:nial-phon}(b), and defined as

\[
\gamma_{j}(\textbf{q})=-\frac{\partial\omega_{j}(\textbf{q})}{\partial V}
\frac{V}{\omega_{j}(\textbf{q})}.
\]

are positive throughout the whole BZ.  There
is no anomalous thermal expansion. Integration of free energies derived 
from phonon spectra gives the thermal expansion shown in 
Fig.~\ref{fig:nial-exp}.
The excellent agreement with experimental data suggests that both 
DFPT phonons and the quasiharmonic approximation are valid in this case.

\subsubsection{Complex structure - PdTi}

For PdTi we calculate phonon spectra for two phases, the low
temperature $B19'$ and the ambient temperature $B19$ phase.  
In both cases, the equilibrium cold curve structure is 
determined for a particular pressure by relaxing internal parameters and 
lattice constants simultaneously\cite{huang:2002b}.  It is then assumed
that the effect  of thermal expansion on lattice parameter ratios and 
internal parameters is equivalent to that of (negative) 
pressure, such that the 
phonon spectrum need only be evaluated once at each volume, and the free 
energy can be interpolated as a function of volume only, with the free 
parameter relaxation done on the cold-curve structure.  The results 
for anisotropic deformation in tungsten (Fig.~\ref{fig:Wtet}) suggest that 
even if the effects of pressure and thermal expansion on free 
parameters are not equivalent, the consequent change in thermal 
free energy due to non-volumetric strains will be small.

\begin{figure}
  \includegraphics[scale=0.4]{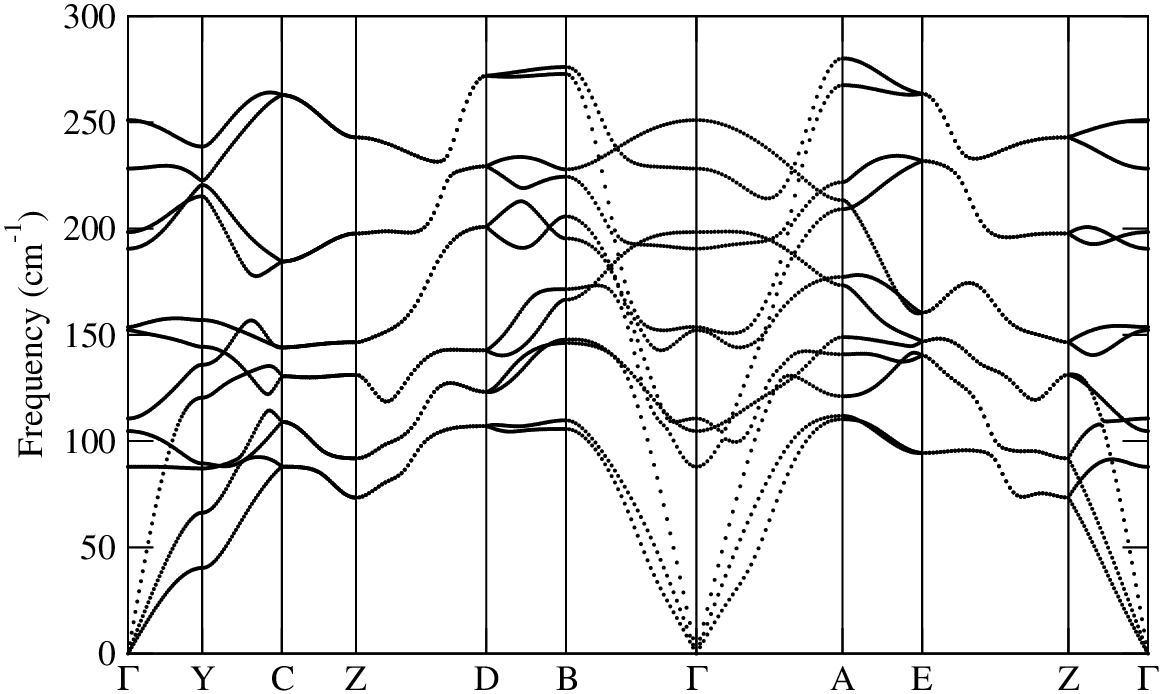}
  \includegraphics[scale=0.4]{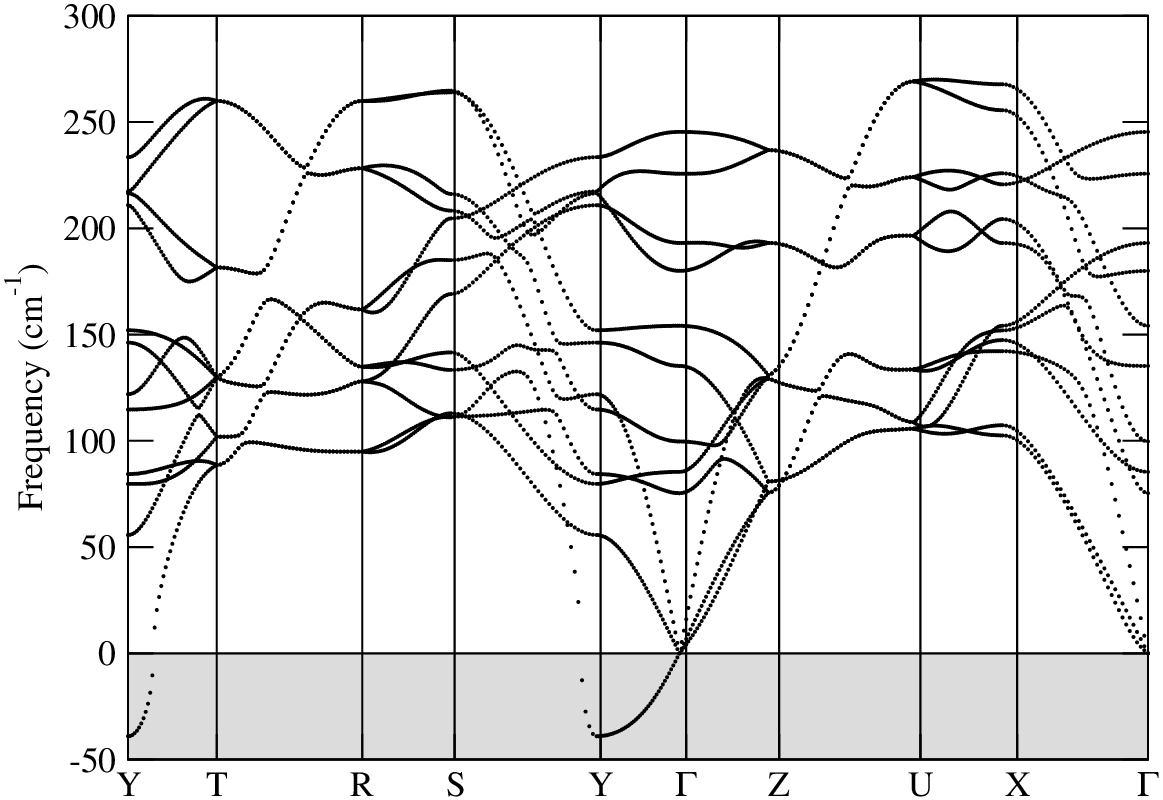}
  \caption { \label{fig:pdti-phon}
  Phonon spectra calculated for (a) $B19'$ PdTi (b) $B19$ PdTi. $B19$ has many soft
  phonon modes which characterize the dynamical instability (grey region, 
these frequencies are actually imaginary). Calculations
  ultrasoft pseudopotentials\cite{rappe} $12\times8 \times 8$ mesh and 30Ry cutoff.
}
\end{figure}

\begin{figure}
  \includegraphics[scale=0.5]{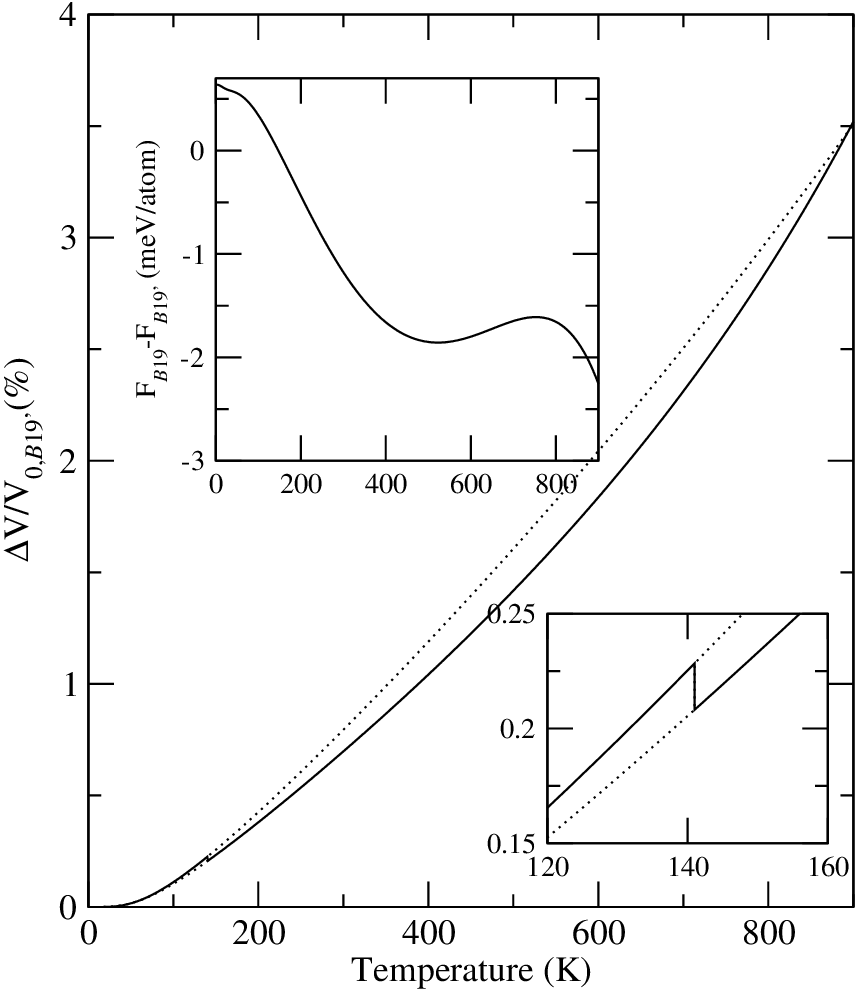}
  \caption {\label{fig:pdti-thermal}
  Calculated percentage thermal expansion for PdTi, 
  including the $B19$-$B19'$ phase transition 
  (expanded scale in lower insert). 
  The dotted lines shows the calculated 
  thermal expansion for the thermodynamically unstable phase 
  (B19 at low T, B19' at high T), assuming 
  that all modes remain harmonic.  The reference volume is the
  T=0 volume for B19': $30.64\AA^3$ which is indistinguishable 
  from the equilibrium for B19: $30.64\AA^3$ and  about 0.34\%
  larger than the cold curve minimum\cite{huang:2002b} 
  which excludes zero-point vibrations.  The upper insert shows the 
  free energy difference between the two phases as a function of 
  temperature, with the phase transition at 140K. This is very 
  much larger than the estimate taken from the Landau barrier height
\protect{\cite{huang:2002b}}}
\end{figure}

The $B19'$ phase has stable phonons throughout the entire BZ 
(Fig.~\ref{fig:pdti-phon}(a)) 
but the $B19$ phase (Fig.~\ref{fig:pdti-phon}(b)) 
has some imaginary frequency modes: 
we assume these contribute to the free energy like free 
particles\cite{dswift}. 
 Assuming the phonons in the $B19'$ phase to remain
harmonic up to and above the phase transition, 
this enables us to calculate the
relative free energies of the two phases, and determine for each
temperature not only the lattice parameters but also which is the
stable phase.  This gives rise to the thermal expansion shown in
Fig.~\ref{fig:pdti-thermal} with a discontinuity marking the
phase transition.  

The calculated phase transition temperature for the $B19$-$B19'$ phase, on
the basis of equal quasiharmonic free energies, is 140K.  This can be
compared with another estimate based on treating the $B19$ phase as
a barrier between equivalent $B19'$ variants, which barrier 
(about 0.0007eV/atom)
can continually be crossed when the temperature reaches 
$\Delta E/k_B=9K$  in the higher temperature phase.  The discrepancy 
between these arises because one concentrates on dynamics of  
a particular soft mode, while the other considers the static 
average over all modes: experimentally, the lower of the two is expected.

\section{Discussion}

We have calculated the thermal expansion properties of W, NiAl and
PdTi using the quasiharmonic approximation and DFPT finding that
agreement with experiment is excellent.  This shows that the harmonic 
phonon free energy is by far the dominant effect in thermal expansion, 
and that other effects such as thermal defects, electron free energy,
coupling of phonons to lattice parameters, and
anharmonic effects due to finite atomic displacements can safely be 
neglected.

 These three materials are
chosen to represent represent an increasingly challenging test to this
methodology. The accurate thermal expansion in tungsten was expected
on the basis of previous work\cite{quong,xie}, and allowed for the
demonstration that anisotropic strain has only a second order effect
on the free energy.  Similarly the results on NiAl show that the method 
can be applied to alloys just as effectively as with elements and showed 
that the presence of phonon anomalies also has little effect on thermal 
expansion.

Without these two results, the PdTi calculation would have been
intractable, but armed with the knowledge that the anisotropic
temperature effect can be treated as a pressure effect on the cold
curve, allowing the vibrational partition function to be treated as a
function of volume only, calculation and minimizations of the full
$F(B_{ij}, {\bf u_k},T)$ becomes tractable.  In addition, 
implementing the previously-described treatment of soft phonons\cite{dswift,drummond} allow us to predict the thermal expansion of both $B19'$ 
and the dynamically stabilized
$B19$ PdTi crystal, in addition to estimating the phase transition temperature
for $B19$-$B19'$.  To our knowledge, these quantities have yet to be measured 
experimentally and such measurement will provide a sensitive test to our 
methods.

In sum, this paper represents a significant theoretical advance in the 
type of materials whose thermal expansion can be calculated from {\it ab initio}
simulation.

This work was carried out under grant AFOSR/MURI F49620-98-1-0433.
The calculations were performed on the SGI Origin 2k/3k at ARL MSRC.
GJA would also like to thank the Fulbright Foundation for support.

%%%%% use bibtex %%%%%

\newpage

%%%%%
%%%%% Fig. 1
%%%%%

%%%%%
%%%%% Fig. 2
%%%%%

%%%%%
%%%%% Fig. 3
%%%%%

%%%%%
%%%%% Fig. 4
%%%%%

%%%%%
%%%%% Fig. 5
%%%%%

%%%%%
%%%%% Fig. 6
%%%%%

%%%%%
%%%%% Fig. 7
%%%%%

%%%%%
%%%%% Fig. 8
%%%%%

%%%%%
%%%%% Fig. 9
%%%%%

\begin{thebibliography}{10}
\bibitem{foot} There is an 
alternate possibility, that thermal expansion is equivalent to 
isotropic rescaling the
lattice constants, however in general this leads to unphysical 
non-zero stresses on the cold curve and was rejected.
\end{thebibliography}
\end{document}